\documentclass{appolb1}
\usepackage{amsfonts}                   
\usepackage{graphicx,color}		

\begin{document}
\newtheorem{theorem}{{\sc Theorem}}[section]
\newtheorem{lemma}{{\sc Lemma}}[section]
\newtheorem{corollary}{{\sc Corollary}}[section]
\newtheorem{proposition}{{\sc Proposition}}[section]
\newtheorem{remark}{{\sc Remark}}

\renewcommand*{\thetheorem}{\thesection.\arabic{theorem}.}
\renewcommand*{\thelemma}{\thesection.\arabic{lemma}.}
\renewcommand*{\thecorollary}{\thesection.\arabic{corollary}.}
\renewcommand*{\theproposition}{\thesection.\arabic{proposition}.}
\renewcommand*{\theremark}{\thesection.\arabic{remark}.}

\title{On Value at Risk for foreign exchange rates - the copula approach
\thanks{Presented at FENS 2006}}%
\author{Piotr Jaworski
\address{Institute of Mathematics, Warsaw University\\
ul. Banacha 2, 02-097 Warszawa, Poland}}

\maketitle

\begin{abstract}
The aim of this paper is to determine
the  Value at Risk ($VaR$) 
 of the portfolio consisting of  long positions in foreign currencies on an emerging market.
Basing on empirical data we restrict ourselves to
 the case when the tail parts of distributions of logarithmic returns 
of these assets follow the power laws and the lower tail of associated copula $C$
follows the power law of degree 1.

We will illustrate the practical usefulness of this approach by the analysis of the exchange
rates of EUR and CHF at the Polish forex market.

\end{abstract}

\section{Introduction}
The present paper is a continuation of \cite{[J2]}. In the previous paper we dealt with the purely asymptotic
estimations, whereas now our goal is to provide some estimates valid for quite a substantial part of the tail.

We shall deal with the following simple case.
An investor operating on an emerging market, has in his portfolio two currencies which are highly dependent,
for example euros (EUR) and Swiss franks (CHF).
Let 
$R_{1}$ and $R_{2}$  be their rates of returns at the end of the investment.
Let $w_i$ be the part of the capital invested in the $i$-th currency, $w_1+w_2=1$, $w_1, w_2 >0$.
So the final value of the investment equals
\[W_1=W_0 \cdot (1+R), \;\;\; R= w_1 R_1 +  w_2 R_2.\]

Our aim is to estimate the risk of keeping the portfolio.
As a measure of risk we shall consider 
 "Value at Risk" ($VaR$), which last years became one of the most popular measures of 
risk in the "practical" quantitative finance 
(see for example \cite{[B2],[RM],[CM],[CDJV],[MN],[L],[JMP],[P]} 
). Roughly speaking the idea is to determine the biggest amount
one can lose on certain confidence level $\alpha$.

If the distribution functions of $R_1$ and $R_2$ are continuous then,
 for the confidence level $1-\alpha$, $VaR$
is determined by the condition
\[P(W_0-W_1 \leq VaR_{1- \alpha})=1-\alpha .\]
Note that if $Q_\alpha$ denotes the $\alpha$ quantile of the rate of return $R$, then
we can denote $VaR$ in the following way
\[VaR_{1-\alpha}=- W_0 Q_\alpha.\]

We shall based on the Sklar theorem, which elucidates the role that copulas play
in the relationship between multivariate distribution functions and their univariate margins
(see \cite{[N],[Jo],[CLV]}). We describe the joint distribution of rates of return
$R_1$ and $R_2$ with the help of a copula $C$
\[P(R_1 \leq x_1, R_2 \leq x_2)=C(F_1(x_1),  F_2(x_2)),\]
where $F_i$ is a distribution function of $R_i$.
Note, that $C$ is the joint distribution function
of the random variables
$F_1(R_1)$ i $F_2(R_2)$.

We recall that a function
\[C:[0,1]^2 \longrightarrow [0,1],\]
is called a copula if
\[C(0,y)=C(x,0)=0,\;\;\;  C(1,y)=y,\; C(x,1)=x,\]
\[x_1 < x_2, y_1<y_2 \Rightarrow C(x_1,y_1) +C(x_2,y_2)- C(x_1,y_2) -C(x_2,y_1) \geq 0.\]

\section{Basic empirical observations}
\subsection{Copulas}

The copulas of financial returns have a specific property, namely they have
uniform tails. 

We recall that a copula 
 $C$ has a uniform lower tail if for sufficiently small 
$q_i$
\[ C(q_1,q_2) \approx L(q_1,q_2),\]
where $L$ is a nonzero function, which is positive 
homogeneous of degree 1 (compare \cite{[E],[J],[J1],[J2],[J3]})
\[ \forall t \geq 0\;\;\;  L(tq_1,tq_2)=tL(q_1,q_2).\]

For the daily exchange rates EUR and CHF in polish z\l oty (PLN) we can observe this
phenomenon even for 10\% part of the tails. On the scatter diagram below (Fig.1)
we plot the ranks of daily returns of EUR and CHF (from January 1995 to April 2006, 2858 returns).
One can observe that there is more points at the lower and upper corners than average.
At the second figure we enlarge the lower corner.\\

\noindent
\includegraphics*[height=7cm,width=12cm]{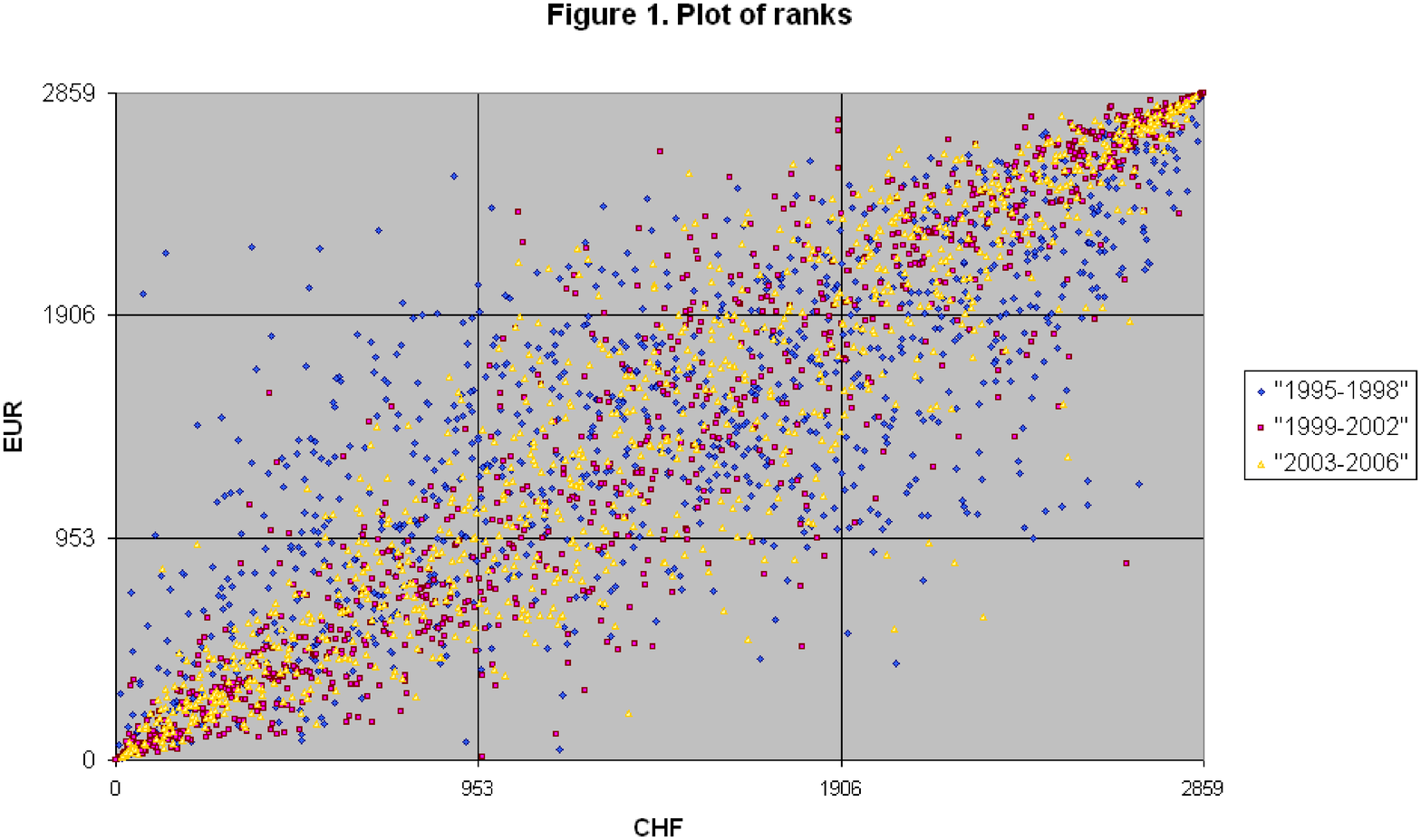}

\noindent
\includegraphics*[height=7cm,width=12cm]{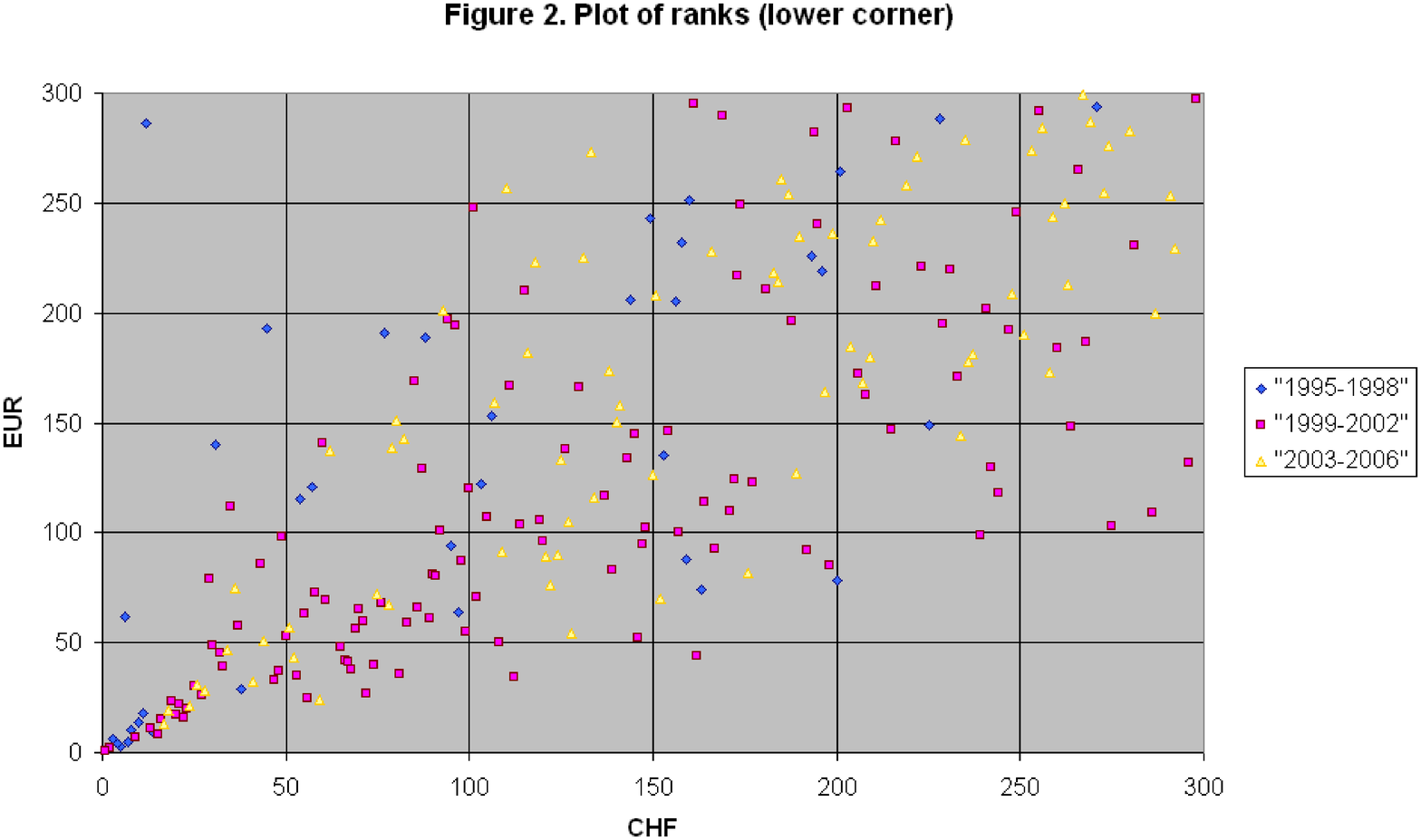}

\pagebreak

To check the homogeneity of the lower tail of the copula we count the number of pairs of ranks
in squares having the origin as a lower vertex
\[ W(n)=\sharp \{ (x_i,y_i)\: : \: x_i \leq n,\, y_i \leq n \}\]
and number of pairs in these squares under and over the diagonal
\[ W_+(n)=\sharp \{ (x_i,y_i)\: : \: x_i <  y_i \leq n \}\]
\[ W_-(n)=\sharp \{ (x_i,y_i)\: : \: y_i < x_i \leq n \}.\]
On figure 3 we show the graphs of these functions. They are close to linear.\\

\noindent
\includegraphics*[height=7cm,width=12cm]{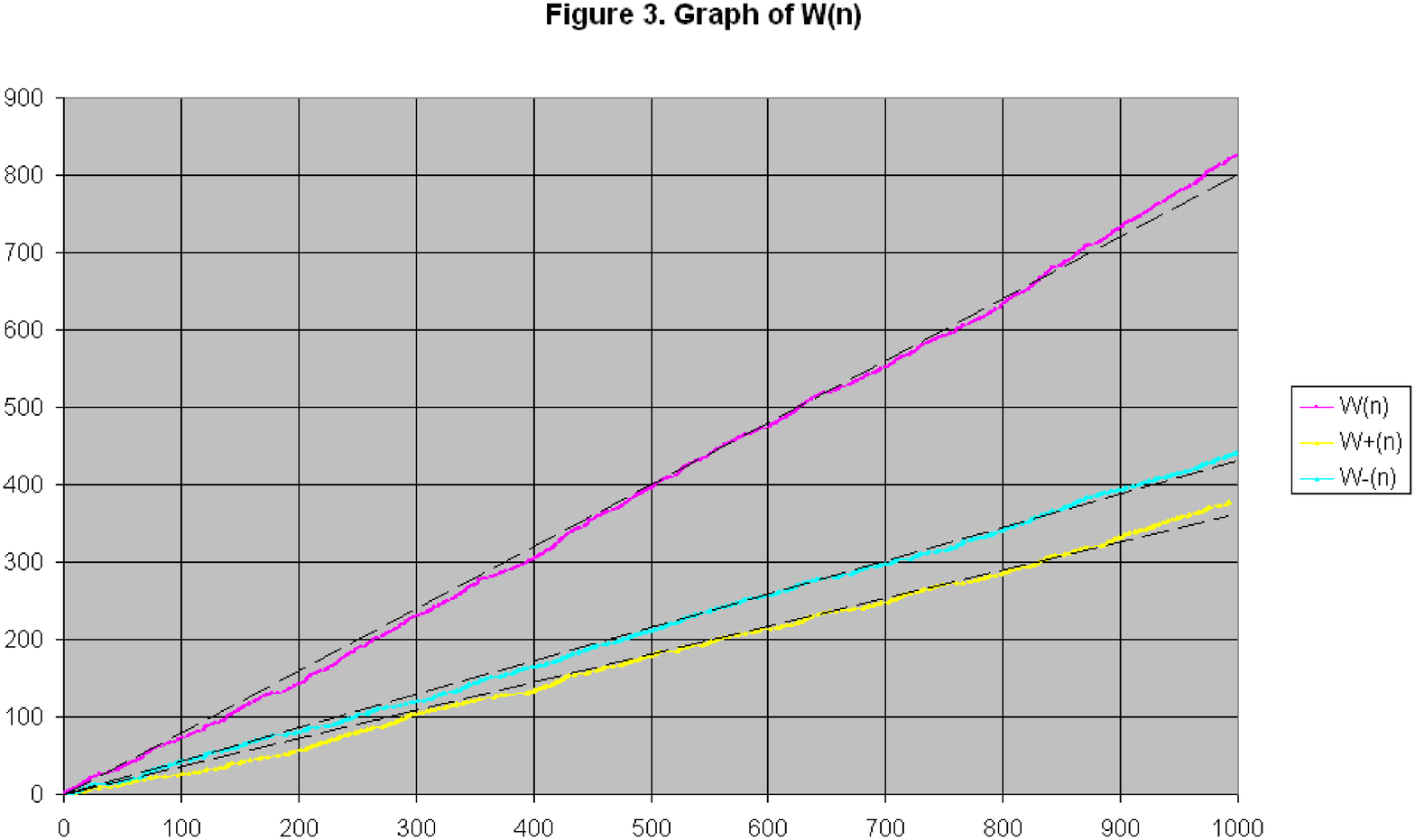}

\subsection{Univariate tails}

The daily log-returns of  exchange rates ($ln(1+R_i)$) have the power-like tails
(compare \cite{[MS]} \S 9.3, \cite{[BP]} \S 2.3.1 or
 \cite{[GMAS],[GGPS],[GGPS1],[DMPV]}).

For sufficiently small $r$ $(-1<r \ll 0)$
\[ F_i(r) \approx a_i \cdot (b_i-ln(1+r))^{-\gamma_i}, \;\;\; i=1,2.\]

For the daily exchange rates EUR and CHF in polish z\l oty (PLN) 
such approximation are valid  even for 10\% part of the lower tails.
On figure 4 we plot the logarithms of minus log-returns against the logarithms of probability.

\noindent
\includegraphics*[height=7cm,width=12cm]{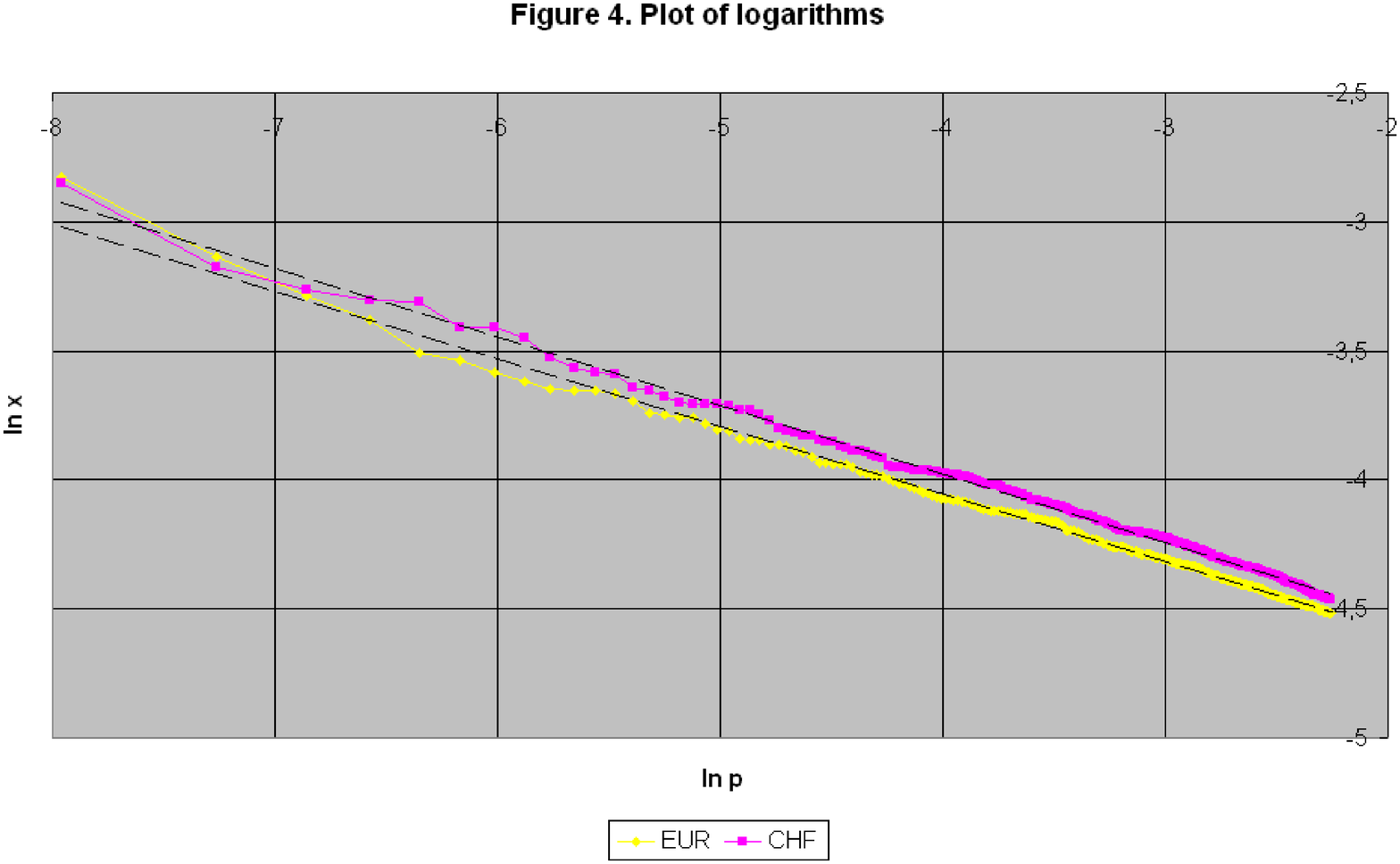}

\section{Main results}
Our aim is to show how to estimate  Value at Risk of the portfolio  ($VaR(W)$) in terms of 
Values at Risk
 of portfolios of the same initial value $W_0$ but consisting only of one currency ($VaR(S_1)$ and $VaR(S_2)$).
The first estimate requires only that the tail part of the copula is homogeneous of degree 1.

\begin{theorem}
If for $q_1, q_2 < \alpha_\ast$ $C(q)=L(q)$, where $L$ is homogeneous of degree 1,
then for $\alpha < \alpha_\ast$
\[VaR_{1-L(1,1)\alpha}(W) \geq w_1 VaR_{1-\alpha}(S_1) + w_2 VaR_{1-\alpha}(S_2).\]
\end{theorem}

The second estimate requires also some properties of lower tails of the marginal distribution.

\begin{theorem}
If for $q_1, q_2 < \alpha_\ast$ $C(q)=L(q)$, where $L$ is homogeneous of degree 1,
and for $-1<t \leq F_i^{-1}(\alpha_\ast)$
\[ F_i(t) = a_i \cdot (b_i-\ln(1+t))^{-\gamma_i},\;\; a_i>0,\;\;  \gamma_i >1,  \;\;\; i=1,2,\]
then for $\alpha < \alpha_\ast$
\[VaR_{1-\alpha}(W) \leq w_1 VaR_{1-\alpha}(S_1) + w_2 VaR_{1-\alpha}(S_2).\]
\end{theorem}

\pagebreak

On figures 5 and 6 we show the plot of the empirical $VaR$ of the portfolio ($w_1=0,4$ EUR, $w_2=0,6$ CHF)
and the  estimates based on the theoretical $VaR$'s for both currencies. We put $W_0=1$.\\

\noindent
\includegraphics*[height=7cm,width=12cm]{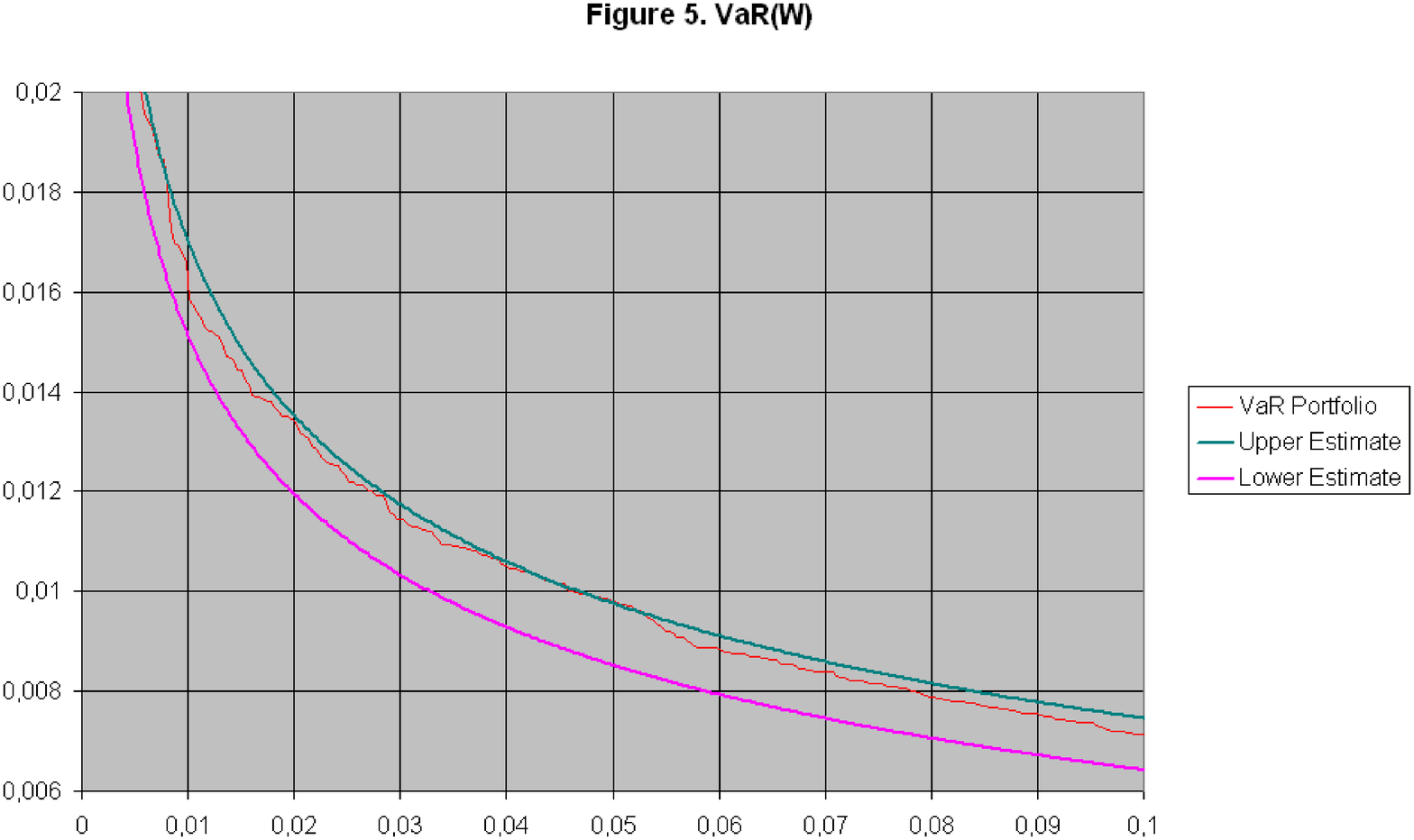}

\noindent
\includegraphics*[height=7cm,width=12cm]{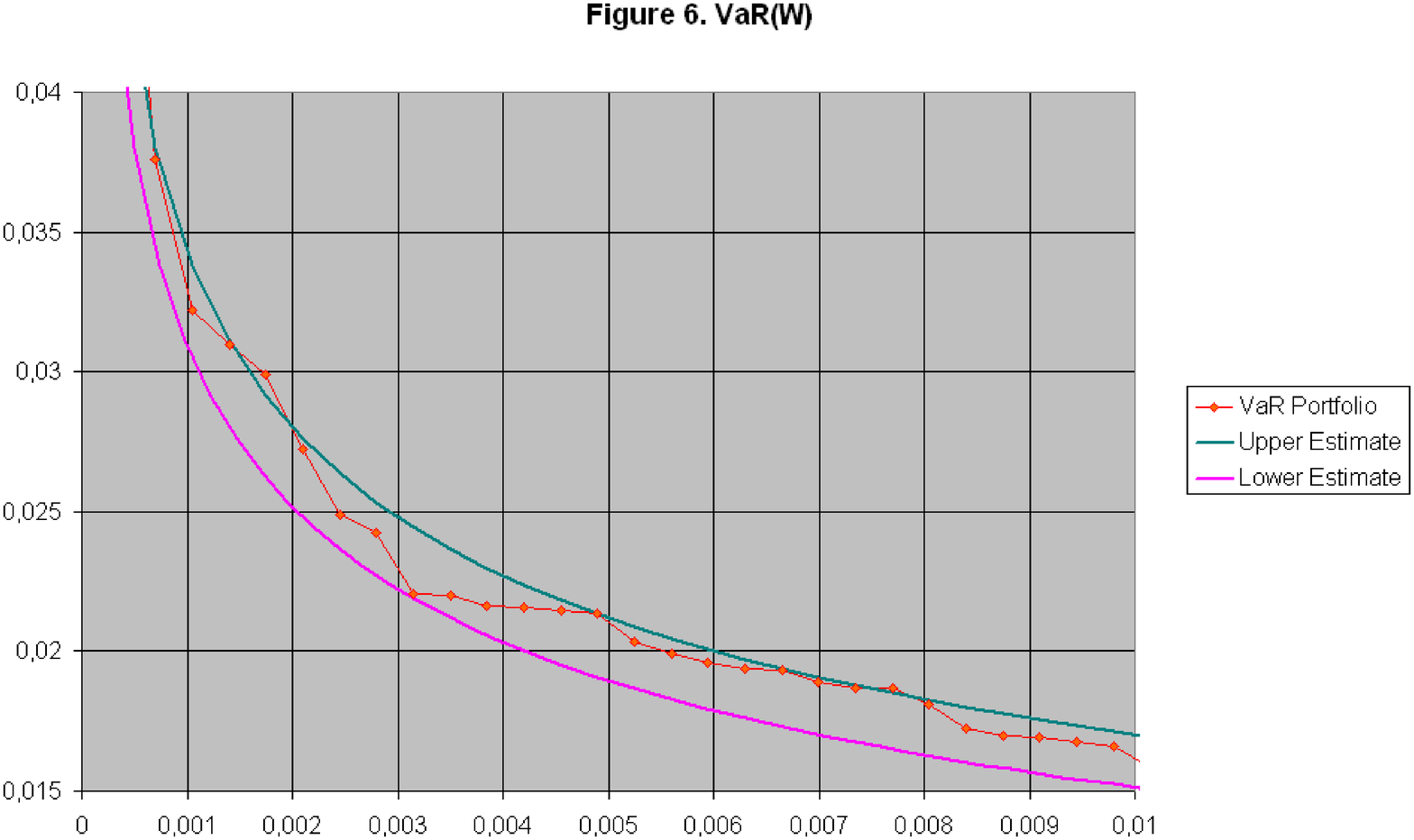}

\pagebreak

\section{Proofs}

\[P(R\leq r)=P(w_1R_1 +w_2R_2 \leq r)= \mu_C( V_r) \approx \mu_L(V_r),\]
where
\[V_r= \{ q: w_1F_1^{-1}(q_1) + w_2 F_2^{-1}(q_2) \leq r \}.\]

Note that the domain $V_r$ is a generalized trapezoid.
\[V_r= \{ q: 0 \leq q_1 \leq q_\ast, \;\; 0 \leq q_2 \leq \varphi_r(q_1) \},\]
where $q_\ast = F_1(\frac{r+w_2}{w_1})$ and $\varphi_r(q_1)= F_2(\frac{r-w_1 F_1^{-1}(q_1)}{w_2})$.

\begin{lemma}
If $rW_0=-w_1 VaR_{1-\alpha}(S_1)-w_2 VaR_{1-\alpha}(S_2)$ then
the square $[0, \alpha] \times [0,\alpha]$ is contained in $V_r$.
\end{lemma}

\noindent
{\em Proof.}
If $q_i \leq \alpha$ then 
\[F_i^{-1}(q_i) \leq F_i^{-1}(\alpha)  = - \frac{ VaR_{1-\alpha}(S_i)}{W_0}.\]
Therefore
\[w_1F_1^{-1}(q_1) + w_2 F_2^{-1}(q_2) \leq -w_1 \frac{VaR_{1-\alpha}(S_1)}{W_0}-w_2 \frac{VaR_{1-\alpha}(S_2)}{W_0}=r.\]

Now we are able to finish the proof of theorem 3.1.\\

\noindent
{\em Proof of theorem 3.1.} (compare \cite{[E1]})\\
Due to the homogeneity we get
\[\mu_C([0, \alpha] \times [0,\alpha])= C(\alpha, \alpha)=L(1,1) \alpha. \] 
Let $rW_0=-w_1 VaR_{1-\alpha}(S_1)-w_2 VaR_{1-\alpha}(S_2)$.
Since the square $[0, \alpha] \times [0,\alpha]$ is contained in $V_r$, we have
\[\mu_C( V_r) \geq L(1,1) \alpha. \]
Therefore the $L(1,1) \alpha $ quantile of $R$ is smaller than $r$. Thus
\[ VaR_{1-L(1,1)\alpha}(W) \geq -rW_0 =w_1 VaR_{1-\alpha}(S_1)+w_2 VaR_{1-\alpha}(S_2).\]
This finishes the proof of theorem 3.1.\\

\begin{lemma}
Let $rW_0=-w_1 VaR_{1-\alpha}(S_1)-w_2 VaR_{1-\alpha}(S_2)$.
If
\[ F_i(t) = a_i \cdot (b_i-\ln(1+t))^{-\gamma_i},\;\; a_i>0,\;\;  \gamma_i >1,  \;\;\; i=1,2,\]
then the function
\[\psi: [0, q_\ast)\longrightarrow [0, +\infty), \;\;\; q_\ast = F_1(\frac{r+w_2}{w_1}),\;\;\;
\psi (q_1)=\frac{q_1}{\varphi_r(q_1)}\]
has the following properties:\\
$\bullet$ $\psi$ is strictly convex and increasing;\\
$\bullet$ $\psi(0)=0$,  $\lim_{q\rightarrow q_\ast^-} \psi(q)= +\infty$, $\psi(\alpha)=1$;\\
$\bullet$ $\lim_{q\rightarrow 0^+} \psi'(q)= F_2((r+w_1)w_2^{-1})^{-1}$, $\lim_{q\rightarrow q_\ast^-} \psi'(q)= +\infty$.
\end{lemma}

\noindent
{\em Proof.}
We have
\[ \psi(q_1)=\frac{q_1} {F_2(\frac{r-w_1 F_1^{-1}(q_1)}{w_2})}=\]
\[=
a_2^{-1} q_1 
\big( b_2 +\ln(w_2) -\ln(1+r -w_1 \exp(b_1 - \big( \frac{q_1}{a_1} \big)^{\frac{-1}{\gamma_1}})\big)^{\gamma_2}.\]
Hence
\[ \psi(0)=0,\;\;\;  \psi(q_\ast^-)=\frac{q_\ast}{F_2(-1)}=+\infty, \;\;\;  \psi(\alpha)=\frac{\alpha}{\alpha}=1.\]
Furthermore
\[ \psi'(q_1)= a_2^{-1} 
 \big( b_2 +\ln(w_2) -ln(1+r -w_1 \exp(b_1 - \big( \frac{q_1}{a_1} \big)^{\frac{-1}{\gamma_1}})\big)^{\gamma_2}
+\]
\[+
a_2^{-1}\gamma_2 \big( b_2 +\ln(w_2) -\ln(1+r -w_1 exp(b_1 - \big( \frac{q_1}{a_1} \big)^{\frac{-1}{\gamma_1}})\big)^{\gamma_2-1}
\times\]
\[\times
\frac{w_1 \exp(b_1 - \big( \frac{q_1}{a_1} \big)^{\frac{-1}{\gamma_1}})}
{1+r -w_1 \exp(b_1 - \big( \frac{q_1}{a_1} \big)^{\frac{-1}{\gamma_1}})}
\frac{1}{\gamma_1} \big( \frac{q_1}{a_1} \big)^{\frac{-1}{\gamma_1}}.\]
Hence
\[ \psi'(0^+)=\frac{1}{F_2((r+w_1)w_2^{-1})},\;\;\;  \psi(q_\ast^-)=\frac{q_\ast}{F_2(-1)}=+\infty.\]
Moreover the first derivative is always positive, hence $\psi$ is strictly increasing.
Also the second derivative is always positive (hence $\psi$ is strictly convex). Indeed:
the second component of the first derivative  is a product of four positive factors,
from which only the last one ($q_1^{-1/\gamma}$) has negative derivative but it is reduced 
by positive derivative of the first component.
\[ \psi''(q_1)= 
 a_2^{-1}\gamma_2 \big( b_2 +\ln(w_2) -\ln(1+r -w_1 exp(b_1 - \big( \frac{q_1}{a_1} \big)^{\frac{-1}{\gamma_1}})\big)^{\gamma_2-1}
\times\]
\[\times
\frac{w_1 \exp(b_1 - \big( \frac{q_1}{a_1} \big)^{\frac{-1}{\gamma_1}})}
{1+r -w_1 \exp(b_1 - \big( \frac{q_1}{a_1} \big)^{\frac{-1}{\gamma_1}})}
\frac{1}{\gamma_1} \big( \frac{q_1}{a_1} \big)^{\frac{-1}{\gamma_1}}q_1^{-1} +\]
\[+ \dots + \dots +\dots +\]
\[+a_2^{-1}\gamma_2 \big( b_2 +\ln(w_2) -\ln(1+r -w_1 exp(b_1 - \big( \frac{q_1}{a_1} \big)^{\frac{-1}{\gamma_1}})\big)^{\gamma_2-1}
\times\]
\[\times
\frac{w_1 \exp(b_1 - \big( \frac{q_1}{a_1} \big)^{\frac{-1}{\gamma_1}})}
{1+r -w_1 \exp(b_1 - \big( \frac{q_1}{a_1} \big)^{\frac{-1}{\gamma_1}})}
\frac{1}{\gamma_1} \big( \frac{q_1}{a_1} \big)^{\frac{-1}{\gamma_1}}\frac{(-1)}{\gamma_1 q_1}=\]
\[= \dots + (\dots )\times q_1^{-1}\big(1-\frac{1}{\gamma_1}\big).\]
Since $\gamma_1$ is  greater then 1 the final result is positive.\\

\begin{lemma}
Let $rW_0=-w_1 VaR_{1-\alpha}(S_1)-w_2 VaR_{1-\alpha}(S_2)$.
If the function $\psi(q_1)=\frac{q_1}{\varphi_r(q_1)}$ has properties listed in lemma 4.2 then
 $\mu_L(V_r) \leq \alpha$.
\end{lemma}

\noindent
{\em Proof.}
\[ \mu_L(V_r)= \mu_L(\{ q: w_1F_1^{-1}(q_1) + w_2 F_2^{-1}(q_2) \leq r \})=\]
\[=
\mu_L(\{ q: 0 \leq q_2 \leq \varphi_r(q_1), 0\leq q_1 \leq q_\ast\})=\]
\[
=\int_0^{q_\ast} \int_0^{\varphi_r(q_1)} \frac{\partial^2 L}{\partial q_1 \partial q_2}(q_1,q_2) dq_2 dq_1
=\int_0^{q_\ast}  \frac{\partial L}{\partial q_1 }(q_1,\varphi_r(q_1))  dq_1.\]
Since $L$ is homogeneous of degree 1, its first derivative is homogeneous of degree 0. Thus
\[ \mu_L(V_r)= 
\int_0^{q_\ast}  \frac{\partial L}{\partial q_1 }(\frac{q_1}{\varphi_r(q_1)},1)  dq_1=\]
\[=
L(\frac{q_1}{\varphi_r(q_1)},1) \frac{ 1}{ \big( \frac{q_1}{\varphi_r(q_1)} \big)' }  |_0^{q_\ast} +
\int_0^{q_\ast}   L(\frac{q_1}{\varphi_r(q_1)},1) 
\frac{\big( \frac{q_1}{\varphi_r(q_1)} \big)''}{\big( \big( \frac{q_1}{\varphi_r(q_1)} \big)' \big)^2}  dq_1\]
For every copula there is  an upper bound $C(q_1,q_2) \leq \min(q_1,q_2)$ (\cite{[N]}).
Since $L$ coincides with $C$ in the lower corner, the same bound is valid for $L$. Therefore
\[ \mu_L(V_r) \leq 
\int_0^{q_\ast}   \min(\frac{q_1}{\varphi_r(q_1)},1)
\frac{\big( \frac{q_1}{\varphi_r(q_1)} \big)''}{\big( \big( \frac{q_1}{\varphi_r(q_1)} \big)' \big)^2}  dq_1=\]
\[=
\int_0^{\alpha}   \frac{q_1}{\varphi_r(q_1)}
\frac{\big( \frac{q_1}{\varphi_r(q_1)} \big)''}{\big( \big( \frac{q_1}{\varphi_r(q_1)} \big)' \big)^2}  dq_1
+
\int_{\alpha}^{q_\ast}   
\frac{\big( \frac{q_1}{\varphi_r(q_1)} \big)''}{\big( \big( \frac{q_1}{\varphi_r(q_1)} \big)' \big)^2}  dq_1=\]
\[=\big(q_1- 
\frac{ \frac{q_1}{\varphi_r(q_1)} }{ \big( \frac{q_1}{\varphi_r(q_1)} \big)' } \big) |_0^\alpha
+\frac{ -1}{ \big( \frac{q_1}{\varphi_r(q_1)} \big)' }  |_\alpha^{q_\ast}=
\alpha -
\frac{ 1}{ \psi'(\alpha)}
+ \frac{ 1}{ \psi'(\alpha)}= \alpha.\]

To finish the proof of theorem 3.2 one has to observe that if
\[ \mu_C(V_r)=\mu_L(V_r) \leq \alpha,\]
then 
\[ VaR_{1-\alpha}(W) \leq -rW_0 =w_1 VaR_{1-\alpha}(S_1)+w_2 VaR_{1-\alpha}(S_2).\]

\end{document}